\documentclass{article}
\usepackage{authblk}
\usepackage[table]{xcolor}
\usepackage[a4paper, margin=1in]{geometry}
\usepackage[utf8]{inputenc}
\usepackage[T1]{fontenc}
\usepackage{graphicx}
\usepackage{multirow}
\usepackage{rotating}
\usepackage[hyphens]{url}
\usepackage{booktabs}
\PassOptionsToPackage{xcolor}{table}
\usepackage{array}
\newcolumntype{P}[1]{>{\centering\arraybackslash}p{#1}}
\usepackage{calrsfs}
\usepackage{amsmath,amssymb,amsfonts}
\usepackage{verbatim}
\usepackage{xcolor}
\providecommand{\keywords}[1]{\textbf{\textit{Index terms---}} #1}
\begin{document}

\title{SoK: Cross-border Criminal Investigations and Digital Evidence}

\author[1,2]{Fran Casino}
\author[3]{Claudia Pina}
\author[1,4]{Pablo López-Aguilar}
\author[1]{Edgar Batista}
\author[1]{Agusti Solanas}
\author[2,5]{Constantinos Patsakis}

 \affil[1]{Department of Computer Engineering and Mathematics, Universitat Rovira i Virgili}
  \affil[2]{Information Management Systems Institute, Athena Research Center, Artemidos 6, Marousi 15125, Greece}
   \affil[3]{European Judicial Cybercrime Network, Eurojust, The Hague, Netherlands}
\affil[4]{Anti-Phishing Working Group - Europe, Av. Diagonal 621–629, 08028 Barcelona, Spain}
  \affil[5]{Department of Informatics, University of Piraeus, 80 Karaoli \& Dimitriou str., 18534 Piraeus, Greece}

\date{}

\maketitle

\abstract{Digital evidence underpin the majority of crimes as their analysis is an integral part of almost every criminal investigation. Even if we temporarily disregard the numerous challenges in the collection and analysis of digital evidence, the exchange of the evidence among the different stakeholders has many thorny issues. Of specific interest are cross-border criminal investigations as the complexity is significantly high due to the heterogeneity of legal frameworks which beyond time bottlenecks can also become prohibiting. The aim of this article is to analyse the current state of practice of cross-border investigations considering the efficacy of current collaboration protocols along with the challenges and drawbacks to be overcome. Further to performing a legally-oriented research treatise, we recall all the challenges raised in the literature and discuss them from a more practical yet global perspective. Thus, this article paves the way to enabling practitioners and stakeholders to leverage horizontal strategies to fill in the identified gaps timely and accurately. }


\section{Introduction}

Understanding the evolution of information and communication technologies without cross-border data flows and ubiquitous systems is impossible. Nevertheless, the opportunities that such evolution brings to all levels of society come with unprecedented challenges in the context of criminal prosecution in cyberspace. Not only the amount of criminal investigations is increasing, but the border-less nature of the Internet adds humongous complexity to such procedures. In addition to the technical challenges of such investigations, the collaboration amongst different organisations is crucial, yet jurisdictional issues further impede it. Thus, the investigation and prosecution of crimes that extend beyond national boundaries is a problem that requires effective measures.

One of the main issues of cross-border investigations is the collection and exchange of electronic evidence, which is often located in multiple countries, requiring external access to it. While being a priority for most countries, there are many unsolved issues due to the different regulatory frameworks of each country, which hinder collaboration due to, e.g., ethical, legal, or even procedural differences. Moreover, since more than half of all criminal investigations require access to cross-border electronic evidence \cite{52018SC0118}, most investigations require evidence requests to other jurisdictions. In addition to the procedural burden, judicial cooperation processes require weeks or even months to be fulfilled.   

The EU and countries such as the United States of America (US) have recently proposed initiatives to address the challenges related to gathering data in different jurisdictions, intending to prevent and prosecute cybercrime in a timely manner. However, as discussed in the literature, these initiatives may pose additional challenges related with fundamental rights and the rule of law provided in the EU and the countries involved.    

\subsection{Contribution}

This article analyses the current state of the art and practice of cross-border investigation initiatives and extracts the main challenges according to their nature, e.g. technological, procedural, legal, communication, economic. It also contributes to the analysis of strategies to overcome them, and provides discussion of the road ahead in cross-border investigations, including research projects, tools, and the impact of technologies such as blockchain that will require novel, adaptable regulations to deal with the dynamic nature of cybercrime. To the best of our knowledge, this is the first article providing such a thorough analysis of the topic, thus enabling a global perspective of the status of cross-border investigations.

The remainder of this work is organized as follows. Section \ref{sec:methodology} describes the research methodology, providing a descriptive analysis of the retrieved literature. Section \ref{sec:background} presents a background on the main initiatives for cross-border data exchange. Section \ref{sec:challenges} describes the state of the art based on the literature analysed in Section \ref{sec:methodology}, and discusses the current challenges of cross-border data exchange initiatives. Relevant open issues, trends, and further research lines are discussed in Section \ref{sec:discussion}. Finally, the article concludes in Section \ref{sec:conclusions} with some final remarks.

\section{Research Methodology} 
\label{sec:methodology}

Our review protocol is based on the five features of Denyer and Tranfield \cite{denyer2009producing} for a systematic literature review. More precisely, the steps are the following: 1) Define the scope of the review 2) Define the research questions 3) Search literature databases 4) Apply inclusion and exclusion criteria, and 5) Synthesise and report the results of the literature analysis.

\subsection{Defining the scope of the review}
A systematic literature review relies on standardised processes for searching, screening, analyzing, and synthesizing the available literature in a systematic, transparent, and reproducible manner, thus assisting in the development of policy and decision-making \cite{tranfield_towards_2003}. Systematic reviews help building a reliable knowledge base by aggregating information from a wide range of relevant studies \cite{tranfield_towards_2003}.

This article focuses on cross-border data exchange initiatives, protocols and solutions, to extract and analyse the current state of practice and the existing challenges to provide a fruitful ground for discussion. Our approach relies on several research questions pertinent to cross-border co-operation, which are aligned to the specific objectives of our article (see Table \ref{tab:resq}). Based on these research questions, we perform a thorough analysis of the available literature and analyse the most well-known protocols and their challenges.

\begin{table*}[hbt!]
\renewcommand{\arraystretch}{1.1}
 \centering
  \caption{Summary of research questions and the corresponding sections devoted to answer them.}
 \scriptsize
 \rowcolors{2}{gray!25}{white}
 \begin{tabular}{p{0.3\linewidth}p{0.5\linewidth}c}
\toprule \textbf{Research Question} & \multicolumn{1}{c}{\textbf{Objective}} & \textbf{Discussion}\\
 \midrule
 Which are the current tools, procedures, and protocols for cross-border evidence exchange amongst European countries/jurisdictions?  & The objective is to summarise the current instruments used between Europe and other countries to leverage cross-border investigations & Section \ref{sec:background} \\
 Which are the main challenges related to cross-border investigations? & The purpose is to collect and summarise the main challenges found in the literature and discuss them  & Section \ref{sec:challenges} \\
 Are current practices efficient enough to counter the sophistication of cybercrime? & The aim of this question is to understand whether current instruments and protocols are sufficient to efficiently fight cybercrime & Sections \ref{sec:challenges} \& \ref{sec:discussion} \\
 What technologies or strategies can be used to deal with the identified challenges? & According to the knowledge extracted from the literature, our plan is to identify the pain points of the actual state of practice to provide fruitful strategies against them. & Section  \ref{sec:discussion} \\
 \bottomrule
 \end{tabular}

 \scriptsize
 \label{tab:resq}%
\end{table*}%

\subsection{Search strategy}
Since we aim to tackle recent challenges of current practice and the impact of novel frameworks in cross-border co-operations, we focused on the last five years, to give an up-to-date view of the current status of the matter.  To this end, we performed a systematic literature search considering papers published between 2016 and 2022 (as of January). Scopus and Web of Science (WoS) were used to locate all scientific-related literature \cite{pranckute2021web}.

We queried Scopus and WoS using the following query: 
\begin{center}\texttt{TITLE-ABS-KEY ( (international  OR  cross-border  OR  cross  AND  border)  AND  investigation  AND  (crime  OR  criminal) )}
\end{center}
It is worth noting that the first bulk search query yielded 379 results. Database's refinement features were used (fine-tuning of results following the context of specific articles, papers, subject area etc.). When a study's abstract was unavailable, the full article was retrieved and evaluated for relevance.

Due to the broad selection of articles, we discovered additional studies using the so-called backward and forward snowball effect, which involved searching the references of articles and reports for additional citations \cite{vom2015standing}. For instance, additional grey literature was discovered by manually searching the reference lists in several reports, since several relevant sources are only present in the form of, e.g. technical reports and guidelines in e.g. official Eurojust, Europol, and international cooperation websites. Following our methodology a total of 442 sources were initially selected (combining research and grey literature).

\subsection{Inclusion and exclusion criteria}

We evaluated the eligibility of the retrieved literature based on a set of inclusion/exclusion criteria. Initially, we excluded all non-English written papers. The next step was the screening of the retrieved papers (title and abstract reading). For the remaining articles, we performed a full reading. It is worth noting that a notable amount papers were excluded during the last two steps (Title/Abstract screening and full paper reading). Our exclusion criteria aimed at fulfilling the scope of the article, thus, we only included articles analysing current cross-border co-operation protocols and methods from a critical perspective, discussing challenges and/or ways to overcome them.

\begin{figure*}[th]
    \centering
    \includegraphics[width=0.8\textwidth]{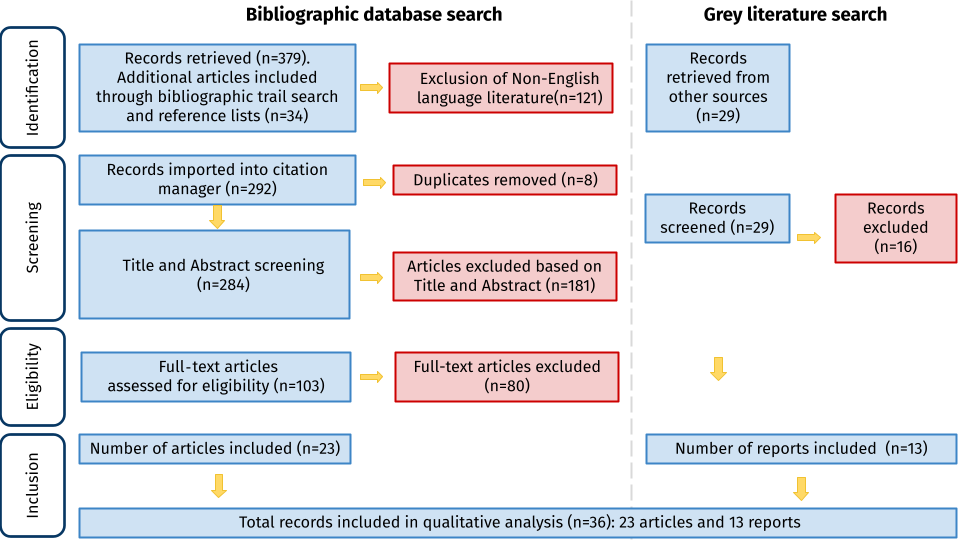}
    \caption{Flowchart of the search strategy.}
    \label{fig:systematic_review}
\end{figure*}

After collecting all the relevant sources and applying our methodology, 103 research articles passed the title and abstract screening. From these, 80 were discarded after a full review, leaving 23 research articles in the scope of our article, which were complemented with 13 sources selected from grey literature. Note that the main focus of the research methodology described in this section is to identify the current challenges of cross-border co-operation and thus, the extracted literature is analysed in Section \ref{sec:challenges}. A summary of the different steps of the bibliographic analysis is depicted in Figure \ref{fig:systematic_review}.

\subsection{Analysis and reporting}

The thematic content analysis enables the descriptive presentation of qualitative data and, therefore, helps researchers identifying, analysing, and interpreting patterns of meaning (or "themes") within qualitative data \cite{Elo2008}. We have adopted a thematic content analysis approach for deriving research areas and common themes from the eligible literature. Due to the nature of some of the reviewed literature (e.g., law-related articles, as well as articles not following well-established sectioning criteria) we combined a qualitative analysis software for the thematic content analysis of the selected literature (MAXQDA2020) with the classical screening and full text reading procedure. Moreover, findings were peer-reviewed by the authors.
Next, we used various ways to synthesise the available literature to report the results of our study in a sound and comprehensive manner. For example, we present the main contributions of each article according to the subset of cross-border protocols analysed, and we extract their challenges in a global manner to derive further discussion.

\section{Main Instruments for Cross-border Investigations in Europe}
\label{sec:background}

Created to address, with due respect for human rights, the legal challenges in criminal justice that emerged from the evolution of technology and telecommunications, the Budapest Convention on Cybercrime of the Council of Europe is the most relevant international instrument on Cybercrime and Digital Evidence.
Opened for signature in 2001, with currently 66 Parties spread around the world, its scope of application is not restricted to the borders of Europe. It aims to create a global framework on cybercrime and digital evidence among practitioners from a very diverse array of jurisdictions, facilitating international cooperation in criminal cases, with substantive and procedural provisions. This legal framework includes provisions for collecting digital evidence in emergencies, directly from service providers, with extra-territorial powers and on international cooperation.
More than 20 years have passed since the drafting of the Budapest Convention, but its criminal substantive aspects, technology neutral in their provisions, remain fully updated. However, in relation to the provisions that support the operational work of Law Enforcement and Judicial Authorities, in view of new introduced technologies such as Cloud Computing and its impact in territoriality and jurisdiction, specific solutions were needed and brought forth by a second Additional Protocol to the Budapest Convention, approved in November 2021\cite{tcy_convention}. 

In response to the identified challenges related to cross border gathering and sharing of digital evidence, the recently approved second Additional Protocol presents new provisions on disclosure of domain name registration information, direct co-operation with service providers for subscriber information, effective means to obtain subscriber information and traffic data, immediate co-operation in emergencies, and a specific provision on Joint Investigations Teams(JITs)\footnote{JITs are a tool in international cooperation in criminal matters, created by a legal agreement between competent authorities of two or more States for the purpose of carrying out criminal investigations, established for a fixed period, usually 12 to 24 months, as needed to conclude the investigation.}. The text was be opened for signature in Strasbourg on the 12th May 2022.
Other very relevant EU legal instruments for lawful collection of electronic information in cross-border investigations are the Mutual Legal Assistance Treaties (MLATs), and the European Investigation Order (EIO),  which replaced MLATs in the context of a subset of participating EU members (EU members except for Denmark and Ireland \cite{ejnjit}). These co-operation instruments rely on the independent judicial scrutiny of the competent authorities in the different countries to guarantee that the corresponding investigation requests and retrieved information are lawfully obtained during investigation processes.

The EIO aims to speed up the co-operation by extending the principle of mutual recognition in evidence gathering. Thus, EU participating member states and their corresponding judicial authorities are entrusted with the task of checking the legitimate grounds to either refuse or execute an EIO. An interesting feature of the EIO Directive is that, in conformity with the EU Charter on Human Rights art 47, allows for the Defense, as well as the victim's lawyer, to request a Court to issue an EIO to obtain digital evidence. This possibility enables lawyers, in equal arms with Prosecution Services, to seek access to the electronic data before it is deleted by requesting the issue of an EIO within the framework of applicable rights of suspects and victims, in conformity with the national criminal procedure or directly in the competent court of the issuing state \cite{stefan2020jud}.

The MLAT process is the most used international co-operation protocol (i.e., MLAT also covers cases in which some of the EU members that want to co-operate are not bound by the EIO Directive). Thus, an MLAT is used to request data residing in countries such as Denmark, Ireland, as well as non-EU countries such as the US or Japan. The main issue with the MLAT requests is that such a procedure was designed before the consolidation of the Cloud as the primary storage platform of most decentralised services on the Internet. Hence, due to this paradigm shift, the increase of cyberthreats that required cross-border co-operation hindered the efficacy of MLATs. As a consequence, the MLAT is currently regarded as an insufficient method to cope with actual needs due to its slowness and the resources that it requires. 

To reduce the burden and speed up the acquisition of electronic data that law enforcement and judicial authorities need for investigating and successfully prosecuting criminals and serious crimes such as terrorism, the EU Commission created the E-evidence initiative, which consists of two main tools, the European Production Order (EPROD) and the European Preservation order (EPRES). The EPROD allows a judicial authority in one Member State to obtain electronic evidence directly from a service provider or its legal representative (thus, entails the creation of such figure in each corresponding service provider) in another Member State \cite{stefan2018cross}. The EPROD imposes a very strict response time (within 6 hours in case of emergencies and to a maximum of 10 days, compared to 120 days in the case of EIO and an average of 10 months for MLAT). The EPRES allows a judicial authority in one Member State to request that a service provider or its legal representative in another Member State preserves specific data given a subsequent request to produce this data by using either an EPROD or an EIO. A parallel instrument with a similar aim was created in the US, namely the Clarifying Lawful Overseas Use of Data (CLOUD) Act \footnote{https://www.justice.gov/dag/cloudact}. One of the most relevant aspects of the CLOUD Act and the E-evidence initiatives is their impact on the actual landscape since most technology corporations are based in the US and EU. Therefore, since both initiatives deviate from the principle by which the physical location in which data are stored determines jurisdiction, and both determine that in specific cases, law enforcement officers should be able to directly access a provider's data under their corresponding jurisdiction without needing an MLAT \cite{abraha2020regulating}, their application could change the cross-border investigation paradigm.   

As noted in the literature, a series of questions are raised as to the E-evidence, and the CLOUD Act’s compatibility with current legal frameworks in relation to privacy, human rights, and the necessity and proportionality principles of the requests made in the context of  cross-border investigations \cite{mulligan2018cross,stefan2018cross}. A clear example of the complexity of the challenges in relation to the E-evidence initiative is the still ongoing trialogue between the EU Commission, Council of the European Union, and European Parliament for the approval of the E-evidence Package. 
In the case of US-EU co-operation, the E-evidence initiative could require US-based online service providers to grant access to data in their possession. At the same time, the Stored Communications Act (SCA\footnote{(SCA, codified at 18 U.S.C. Chapter 121 §§ 2701–2712)}) forbids the provision of such access, unless there is an executive agreement with the US. On the other side, when US authorities request data stored in the EU, companies may risk breaching the EU General Data Protection Regulation (GDPR) under Article 48, since any judgment or decision of an administrative authority of a third country
requiring a controller or processor to transfer or disclose personal data may only be
recognised or enforceable if it is based on an international agreement, such as an MLAT. Moreover, Article 46 of the GDPR also hinders the execution of data exchange procedures on the European side if there is no mechanism allowing European individuals to have the safeguards and legal remedies comparable to those resulting from the GDPR \cite{gppi2019,stefan2018cross}.

\begin{figure*}[th]
\centering
\includegraphics[trim={0 10cm 0cm 0cm},clip,width=\textwidth]{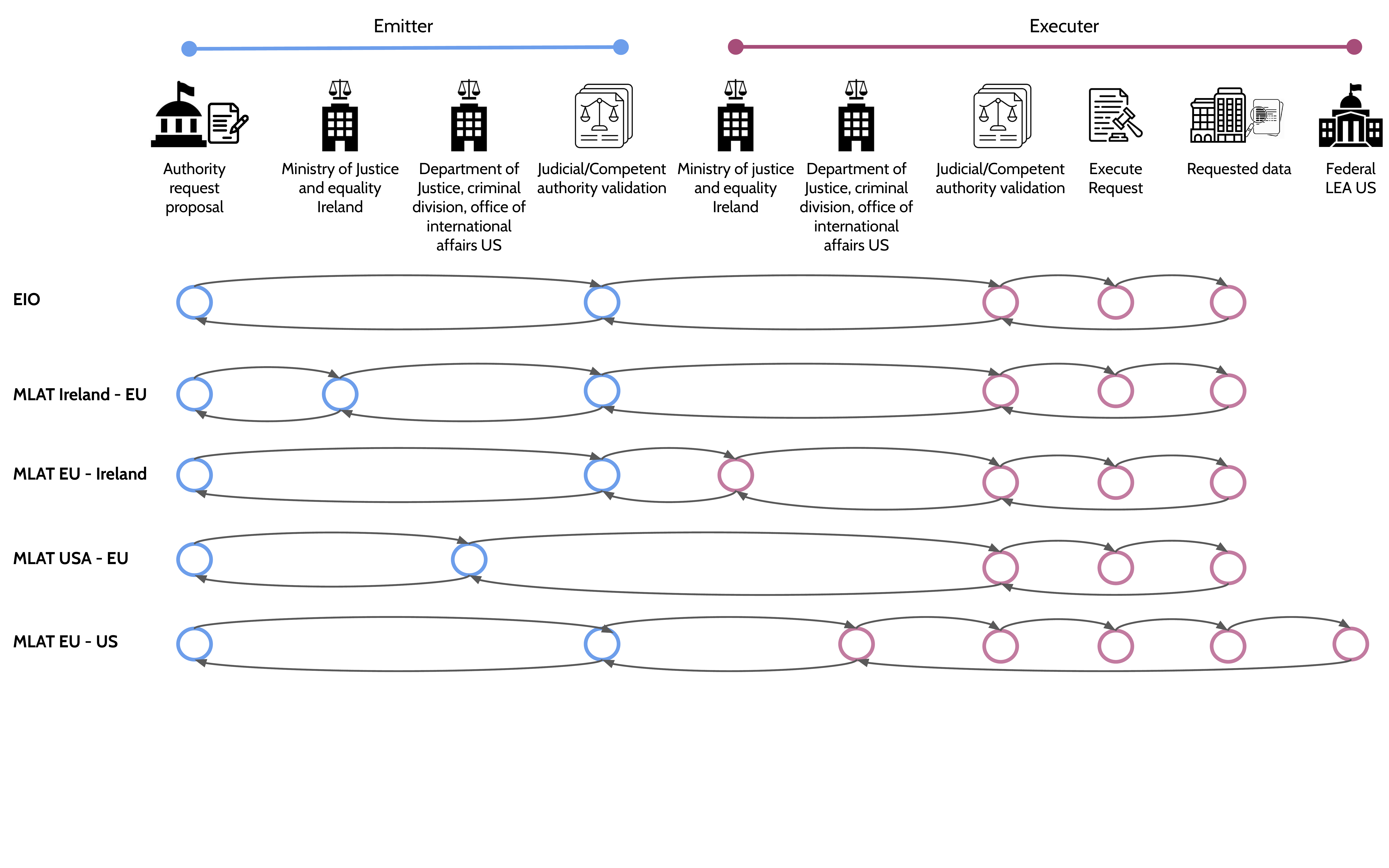}
\caption{Main flow of each collaboration instrument and the corresponding institutions and authorities involved.} 
\label{fig:circles}
\end{figure*}

Table \ref{tab:summary2} summarises the main cross-border investigation instrument and their jurisdictional applicability. Moreover, Figure \ref{fig:circles} shows the different collaboration flows according to each instrument. For a profound analysis of the main co-operation instruments between different countries, we refer the reader to \cite{europol2021sirius,gppi2019,mitsilegas2020cross,stefan2020jud,stefan2018cross}.

\begin{table*}[hbt!]
\renewcommand{\arraystretch}{1.1}
 \centering
  \caption{Main EU legal instruments for channelling cross-border requests for data gathering in criminal proceedings.}
 \scriptsize
 \rowcolors{2}{gray!25}{white}
 \begin{tabular}{cp{4in}}
\toprule
\textbf{Protocol/ co-operation tool} & \textbf{Description and applicability} \\
 \midrule
 CoE Cybercrime convention \footnote{Convention on Cybercrime of 2001 (ETS No. 185)} & 66 countries \footnote{https://www.coe.int/en/web/cybercrime/parties-observers}  \\
EIO\footnote{Directive 2014/41/EU of the European Parliament and of the Council of 3 April 2014 regarding the European Investigation Order in criminal matters, OJ L 130, 1.5.2014} &   EU excluding Ireland and Denmark, since 2014. \\
MLAT within EU\footnote{Mutual Legal Assistance Convention (between Member States of EU)   Council Act of 29 May 2000 establishing in accordance with Article 34 of the Treaty on European Union the Convention on Mutual Assistance in Criminal Matters between the Member States of the European Union, OJ C}    &  Member States of the European Union (special rules apply for Ireland, Norway, Luxembourg, and Iceland),  2000  \\
MLAT between EU and third countries & For example, MLAT with Ireland\footnote{https://revisedacts.lawreform.ie/eli/2008/act/7/revised/en/html} (2008), MLAT EU-Japan \footnote{\url{https://eur-lex.europa.eu/legal-content/EN/TXT/HTML/?uri=CELEX:22010A0212(01)}} (2009), and MLAT with US\footnote{EU-US Agreement on Mutual Legal Assistance (MLA) Agreement of 25 June 2003 on mutual legal assistance between the European Union and the United States of America.} (2003). \\ 
 \bottomrule
 \end{tabular}

 \scriptsize
 \label{tab:summary2}%
\end{table*}%

Beyond the current adopted measures described above, it should be highlighted that by acknowledging the pains and gaps of digital evidence exchange, the international community is making an attempt to harmonise these procedures. Therefore, the proposal for a United Nations Convention on Countering the Use of Information and Communications Technologies for Criminal Purposes is currently being discussed. This discussion, currently being carried out by an Ad Hoc Committee\footnote{Resolution 74/247 adopted by the General Assembly on 27 December 2019} created in December 2019. The Convention, still at an early stage of development, has a focus on State sovereignty and non-intervention. In addition to criminal justice related topics, it aims to cover certain aspects related to Internet governance, cybersecurity, obligations to private sector, and involvement of the International Telecommunication Union (ITU). It is also introducing  new computer-related or cyber-enabled offences such as digital data to mislead users, incitement to subversive activities, terrorism, extremism, drugs, and arms trafficking. The negotiations appear to be very complex to develop  and no date is previewed for their conclusion.  

\section{Literature Review and Challenge Extraction}

\label{sec:challenges}

In this section, we analyse the literature collected following the methodology described in Section \ref{sec:methodology}. For each article, we extract the challenges and group them in a higher level of abstraction, when possible, to provide a more comprehensive perspective of the current state of practice. 

Several articles \cite{abraha2021law,Jerman2019,gppi2019,mitsilegas2020cross,siry2019cloudy,stefan2018cross} recall the issues of MLAT and EIO systems and analyse recent initiatives designed to overcome most of them, namely the CLOUD Act and the E-evidence framework, and highlight the current constraints of this novel proposed legislation and the legal and ethical conflicts across EU and other countries such as the US. 
In \cite{Chauhan2021}, the authors recall the inefficacy of actual protocols such as MLATs for distributed and highly-volatile data in cross-border investigations and highlight the current challenges for collecting evidence from cloud service providers. Similarly, in \cite{kahvedvzic2016cybercrime}, authors discuss the technical issues of cross-border investigations due to the decentralisation of data and the conflicts between EU-US resulting from, e.g. GDPR, which prevents the indiscriminate data request from US to EU providers, recalling the case of Microsoft Ireland. In \cite{shalaginov2020modern}, the authors analyse current cross-border frameworks' issues from a technological and forensics perspective. In \cite{fuster2020cross}, the authors focus on the benefits and drawbacks of the E-evidence initiative. In \cite{kleijssen2017cybercrime}, the authors highlight the current issues towards achieving harmonisation and the difficulties for establishing an equivalent, fluent collaboration between parties using different legislation and definitions, which is crucial to guarantee an effective and timely prosecution. The article presented in \cite{karas2019evaluation} discussed the main issues of EIOs, including an analysis of incoherent definitions affecting the compatibility between legal systems of different countries.  

Some articles focus on tools to complement or substitute the EIO. For instance, \cite{Zaharieva2017} compares the EIO with the Joint Investigation Team (JIT) as tools to ease cross-border investigations. Notably, JIT enables more flexibility to specific investigations regarding the number of authorities implied and their corresponding evidence exchange, yet it requires more coordination and the corresponding agreements between parties. The article presented in \cite{blazic2020} discusses the issues of cross-border access to evidence from the EIO and the E-evidence perspective. Moreover, the author analyses data collected from two surveys among practitioners, showcasing existing obstacles of current frameworks. In \cite{csuri2017}, the authors analyse two tools for judicial co-operation in criminal matters in the EU: the European Investigation Order (EIO) and the proposed European Public
Prosecutor’s Office (EPPO) \cite{epporegister}, showing the lack of standardised mechanisms of the EPPO. Thus, given the considerable differences between the national legal systems, the consistency of the EPPO investigations and their admissibility in court require complex rules for mutual admissibility of evidence that may not be realisable. 

Other authors focus on specific parts of these frameworks. In \cite{Warken2020}, the authors analyse the different data categorisation schemes of current evidence exchange frameworks, namely, EIO, E-evidence, and the CLOUD Act, highlighting the need for harmonising and updating the different definitions and types of data. The latter is critical to ensure that the proper data requests are issued to fit the purpose of the investigation, being compliant with legislation. The authors propose a set of data categories to minimise the incoherence among such frameworks. Similarly, a review of the main challenges to adopting novel directives, especially in terms of data categorisation and the related legal issues, is presented in  \cite{Jerman2020}. The issue of data ownership in distributed systems is analysed in \cite{karagiannis2021digital}, where authors recall the power of disposal as a possible solution to ease and speed up the data collection procedures in distributed systems while recalling the current issues of the EIO in that regard. The article in \cite{stefan2020jud} provides a guideline for the different actors participating in cross-border investigations by analysing the existing frameworks aiming to enable criminal justice co-operation. In \cite{Biasiotti2018}, the authors discuss the difficulties and lack of harmonisation in regards to the evidence categorisation and classification necessary to establish standardised procedures and enhance collaboration between different organisations. In \cite{Barbosa2019}, the authors discuss the applicability of the speciality rule in the EIO that enables the use of evidence gathered in the context of an investigation to be used for other purposes. Notably, despite the ambiguities that could prevent the effective use of such evidence, the EIO does not provide strict prohibitive measures, thus aiming at a path of free movement of evidence in the EU. Since the E-evidence framework is far more restrictive at that level, the EIO is usually preferred, yet such ``freedom" can be misused to circumvent legal prohibitions.

The work presented in \cite{OrtizPradillo2017} discusses the law updates of different EU countries to cope with new technologies. Moreover, it discusses several vague directives of the Cybercrime Convention Committee (T-CY) proposed in April 2013 \footnote{Council of Europe Cybercrime Convention Committee (T-CY): (Draft) Elements of an additional protocol to the Budapest convention on cybercrime regarding cross-border access to data (April 9, 2013).} with particular attention to the exact location of the data to be seized. The authors discuss the desirability of a regulation that stipulates that access to stored computer data should be possible regardless of where the data are located if specific cumulative requirements are met (e.g., the exact location of data is not known, the evidence has to be retrieved in a lawful and proportionate manner, and for specific purposes) to solve ambiguous/uncertain situations, including the cases where anonymising tools (e.g., The Onion Router - TOR) are used. In the context of financial crimes, \cite{pavlidis2021asset} analyse the mutual recognition of freezing and confiscation orders, which work in conjunction with other EU mutual recognition instruments such as the European Arrest Warrants and the EIOs to speed up the enforcement of financial-related orders in all member states. 

In \cite{birdi2021factors}, the authors use a sound methodology to create ten use cases and collect the challenges related to knowledge sharing between organisations across international boundaries. The authors classify such challenges and provide some recommendations to overcome them.  

Several articles discuss specific international collaborations and their impact on liaising with the EU. For instance, \cite{Heusala2018} analyses the benefits and challenges of the bilateral co-operation between Finland and Russia, which is based on the Treaty on Crime Prevention co-operation between the governments of Finland and the Russian Federation (1994). The authors highlight that the practical challenges in such bilateral agreements overlap with those observed in other multi-lateral agreements at the EU level.  
Recent Brexit-related issues are discussed in \cite{loik2020european} along with current mutual recognition instruments, which will have to be redefined to maintain the EU-UK co-operation. 
In the US context, \cite{Currie2017} analyses the extraterritorial enforcement to electronic evidence issues with a particular focus on the Microsoft Ireland case. In \cite{Ghappour2017}, the authors raise the concern of using dubious tools by the US government to collect evidence (with particular focus on the dark web) in foreign countries and their legal implications. In \cite{Arrigg2019}, the authors discuss the issues faced by the judiciary system, prosecution, and other actors involved in foreign and cross-border investigations from the US perspective. They provide some insights on how to overcome and/or minimise them. Similarly, the authors in \cite{modi2017toward} recall the issues regarding the applicability of the suspects' rights when there are conflicts between different jurisdictions, focusing on the US Fifth Amendment. In \cite{mulligan2018cross}, the authors analyse the primary forms of cross-border data sharing with the US, namely, letters rogatory, MLATs, and executive agreements authorised by the CLOUD Act, and highlight the benefits and limitations of the CLOUD Act compared with the other two previous protocols. Similarly, in \cite{abraha2020regulating} the authors examine the impact, the opportunities, and the adequacy of the US CLOUD Act concerning other international governments.


\begin{table*}[hbt!]
\renewcommand{\arraystretch}{1.1}
 \centering
 \caption{High level abstraction and description of the challenges identified in the literature.}
 \scriptsize
 \rowcolors{2}{gray!25}{white}
 \begin{tabular}{p{0.23\linewidth}p{0.73\linewidth}}
\toprule \textbf{Challenge} & \multicolumn{1}{c}{\textbf{Description}} \\
 \midrule
 Data location and individuals' control over their own data &  This challenge recalls the difficulty to establish the exact location(s) and existing copies of each individual's data, who has access to it, and on which grounds. For instance, proposals such as the E-evidence are not clear towards this aspect and introduce uncertainty.\\
  Timely collection, analysis and sharing of evidence &  In this category, we include issues related to the fact that nowadays investigations may require the processing of vast amounts of data, along with its volatile nature. The latter is aggravated due to the inefficacy of MLAT and EIO frameworks,  which may not accommodate the necessary speed or provisions to facilitate evidence collection.  \\
      Lack of harmonisation in rules of admissibility of criminal evidence and prosecution & The lack of clear and common legislation regimes and standards across different states in relation to data retention, the gathering and validity of digital evidence versus the procedural rights of potential suspects, may hinder the applicability of the EIO directive due to jurisdictional constraints (especially in the case of protocols that do not require an independent judiciary validation). This situation, apart from the potential creation of inconsistent prosecution scenarios may be exploited by criminals using jurisdictional arbitrage tactics in cases of remarkable differences among states.\\
  Lack of compatibility between different protocols regarding data categorisation and definitions & 
This category includes the issues related to communication between different jurisdictions or states, in which potential incompatibilities regarding data categorisation and related definitions may arise. The latter could create conflicts when applying the appropriate legal standards and procedures when requesting and processing new evidence, which could entail delays and even affect the validity of evidence in court if the proper procedures were not followed. \\
 Direct cooperation with service providers and equality of opportunities  & This category includes several issues related to the user's side and the equality of legal rights and opportunities. For instance, in the context of novel protocols such as E-evidence, the lack of information and procedural details hinders the refusal of a production or preservation order. More concretely, individuals and private companies' representatives may lack clear indications where and to whom to bring their claims or assess whether the petition satisfies law requirements regarding judicial independence. \\
 Incompatibility conflicts between jurisdictions that may violate procedural rights and safeguards & Non-judiciary mediated orders (i.e., orders directly issued from police or prosecutors such as the ones foreseen in the case of E-evidence and CLOUD Act) lower the standards previously necessary to obtain evidence in cross-border criminal investigations and prosecutions. Note that the information obtained during investigations has to be specific to ensure defence rights in criminal proceedings, including the basis for the request, how the search was done, and how the data was analysed by investigating/prosecuting authorities. Moreover, specific bi-lateral incompatibilities arise between the CLOUD Act and EU (CLOUD Act breaches Article 48 of GDPR), and the E-evidence and US (the E-evidence may require data for which  access is forbidden by the US Stored Communications Act). The latter issues are exacerbated by further incompatibilities between the legislation of different states, which may hinder international co-operation and its judicial robustness. \\
Lack of automated mechanisms to efficiently collect and report requests & Generally, EU member states do not have a unified system for collecting and reporting information related to issued/received cross-border data requests, channels/instruments used, and related outcomes. There are essential transparency deficits regarding how data requests are issued, transmitted and executed by competent national authorities.\\
 Auditability in data collection procedures & This category recalls data collection and management issues during investigations. In this sense, most current mechanisms lack standardised procedures to ensure that data gathering is consistent with national and/or fundamental international rights and the rule of law standards that apply to criminal investigations so that evidence is admissible in court. Moreover, post-investigation management of evidence is not properly tackled in approaches such as the EIO. For instance, EIOs do not deal with the use of outcomes obtained from shared evidence and the possibility to further share them with other parties or use them in other investigations.\\
  Lack of resources related to equipment and training of law enforcement and judicial authorities to support direct co-operation between different jurisdictions & There is a lack of resources and efforts for training personnel to support the investigations both in the technical and the legal aspects, including information about the applicable rules and procedures considering the particularities of different legal systems. \\
  Data Retention issues & Considering the already mentioned needs for digital evidence in criminal investigations the availability of information is crucial to criminal investigations, however, quite often, the required digital evidence is in the possession of telecommunications service providers and therefore, retention of non-content communications data is a very relevant issue to be considered by public authorities.  \\
 \bottomrule
 \end{tabular}
 
 \scriptsize
 \label{tab:challenges_exp}%
\end{table*}%

\begin{table*}[hbt!]
\renewcommand{\arraystretch}{1.1}
 \centering
 \caption{Relation of challenges discussed by each corresponding article. }
 \scriptsize
 \rowcolors{2}{gray!25}{white}
 \begin{tabular}{p{4in}p{2.5in}}
\toprule \textbf{Challenge} & \multicolumn{1}{c}{\textbf{Reference}} \\
 \midrule
  Individuals' control over their own data &   \cite{stefan2018cross,siry2019cloudy,Ghappour2017,shalaginov2020modern,karagiannis2021digital}  \\
   Timely collection, analysis and sharing of evidence &  \cite{stefan2018cross,gppi2019,mitsilegas2020cross,fuster2020cross,abraha2020regulating,Jerman2020,Currie2017,abraha2021law,stefan2020jud,abraha2020regulating,siry2019cloudy,Zaharieva2017,blazic2020,Ghappour2017,csuri2017,Mitsilegas2017,Chauhan2021,mulligan2018cross,kahvedvzic2016cybercrime,pavlidis2021asset,kleijssen2017cybercrime}\\
    Lack of harmonisation in rules of admissibility of criminal evidence and prosecution & \cite{mitsilegas2020cross,OrtizPradillo2017,birdi2021factors,Arrigg2019,Berge2021,kahvedvzic2016cybercrime,modi2017toward,loik2020european,shalaginov2020modern,pavlidis2021asset,karas2019evaluation}	\\
    Lack of compatibility between different protocols regarding data categorisation and definitions & \cite{Warken2020,Heusala2018,Jerman2020,blazic2020,OrtizPradillo2017,birdi2021factors,Biasiotti2018,Arrigg2019} \\
  Direct cooperation with service providers and equality of opportunities &  \cite{stefan2018cross,mitsilegas2020cross,gppi2019,stefan2020jud,blazic2020,Arrigg2019,Ghappour2017,modi2017toward,europol2021sirius,europolchallenges}\\
  Incompatibility conflicts between jurisdictions that may violate procedural rights and safeguards&  \cite{stefan2018cross,mulligan2018cross,mitsilegas2020cross,gppi2019,Jerman2019,Currie2017,abraha2021law,abraha2020regulating,siry2019cloudy,blazic2020,Berge2021,fuster2020cross,Ghappour2017,Barbosa2019,Mitsilegas2017,Heusala2018,birdi2021factors,Arrigg2019,kahvedvzic2016cybercrime,modi2017toward,kleijssen2017cybercrime,karas2019evaluation,karagiannis2021digital} \\
  Lack of automated mechanisms to efficiently collect and report requests  &\cite{mitsilegas2020cross,Chauhan2021,birdi2021factors}	 \\
  Auditability in data collection procedures &  \cite{stefan2018cross,stefan2020jud,Barbosa2019,Chauhan2021,karas2019evaluation}  \\
  Lack of resources related to equipment and training of law-enforcement and judicial authorities to support direct co-operation between different jurisdictions & \cite{stefan2020jud,Jerman2020,blazic2020,Jerman2019,mitsilegas2020cross,Heusala2018,birdi2021factors,pavlidis2021asset,kleijssen2017cybercrime}\\ 
  Data retention issues & \cite{gppi2019,stefan2020jud,europol2021sirius,europolchallenges}\\
 \bottomrule
 \end{tabular}
 
 \scriptsize 
 \label{tab:challenges_rel}%
\end{table*}%

After collecting the challenges from the reviewed literature and comparing them with public information collected from judicial and law enforcement authorities \cite{europol2021sirius}, we have categorised them as seen in Table \ref{tab:challenges_exp}. Thereafter, we mapped each challenge with the corresponding articles where they are discussed in Table \ref{tab:challenges_rel}. We observe that one of the most recalled challenge is the timely collection and sharing of evidence in the context of MLATs and EIO, which is especially relevant in cases where data is highly volatile \cite{casino2022research,europolchallenges,europol2021sirius,kleijssen2017cybercrime}. In this regard, one strategy to enhance the efficiency of MLATs is to provide more descriptive definitions about data requests and to increase transparency of all parties involved in the whole procedure \cite{kleijssen2017cybercrime,gppi2019}. Another strategy proposed in the literature is to deploy specialised personnel and national contact points to establish the necessary agreements between different legal systems \cite{stefan2018cross}. The latter is related to another struggling challenge: the lack of resources for cross-border collaborations, which is critical since the proper training of personnel and the enhancement of their technical skills are crucial to guarantee quality and timely investigations. Notwithstanding, beyond finding qualified personnel, the budget devoted to this goal has to be enough to guarantee not only the engagement of professionals but also to provide them with the proper tools and equipment \cite{birdi2021factors,gppi2019,stefan2018cross}.

The lack of automated mechanisms to speed up cross-border investigations and the auditability of the interactions and collected evidence are two challenges that require the use of tools with verifiability and auditability capabilities \cite{casino2019systematic,casino2022research,javed2022comprehensive}. Moreover, there is a need for defining standardised procedures, even if it is at a bi-lateral level so as to speed up the interactions and increase trust among organisations \cite{gppi2019,mitsilegas2020cross}. There are some examples of tools aiming at easing cross-border collaboration, such as the e-CODEX \cite{ecodexevidnece2e} system, which enables both requests and evidence to be exchanged securely between judicial authorities.
Similarly, the e-Evidence Digital Exchange System (eEDES) \cite{eEDES} aims to establish a secure and decentralised platform in Europe to ease communications and evidence exchange, particularly in the context of EIO and MLAT cross-border investigations. Another relevant measure at the international level is INTERPOL's e-MLA initiative \cite{emla}, which aims to develop a platform for collaboration and Mutual Legal Assistance (MLA) exchanges. A nice addition to these tools could be to leverage the prosecution of criminal activities by using Eurojust's recommendations \cite{eurojustrecomm}, which offer the legal framework to decide, in the case of a criminal activity that is being prosecuted in different member states, which of them is in a better position to undertake an investigation or prosecute specific acts, avoiding duplicated efforts and reducing the investigations overhead. Complementary to the tools, a more in-depth focus on the different modes of collaboration among organisations is critical, since understanding organisational differences and establishing good relationships with policing organisations is critical to enhance the mutual trust \cite{birdi2021factors}. In this regard, one of the current issues is the proper regulation and definition of data types and their possible categorisation to facilitate the efficacy of MLATs, and cross-border co-operation \cite{kleijssen2017cybercrime,Warken2020}. The latter is crucial in financial crime investigations in which the need for more transparent monitoring of virtual assets (including the possible establishments of central bank account registries) and auditable data collection procedures are mandatory to enforce the corresponding required orders \cite{pavlidis2021asset}.  

Although judicial cooperation instruments remain essential mechanisms to obtaining electronic evidence, especially when gathering content data, they are deemed too slow to share electronic evidence effectively.
In order to obtain data more swiftly, keeping up with a constantly evolving digital landscape and a fragmented legal framework, public authorities seek direct cooperation with Online Service Providers (OSPs), a path that is not without difficulties, due to a very diverse array of potentially applicable rules, the need for taking into account the internal procedures of OSPs and consequent uncertainty for public authorities, citizens and the involved private entities. For instance, users, companies or industries may be forced to accept requests by default to avoid possible sanctions. In addition, defendant lawyers lack mechanisms to issue a data collection procedure (this is only foreseen in the EIO framework), so they do not possess the capabilities to request data that could be used as evidence to properly exercise the defendants' rights. Moreover, conflicts may arise in cases where the law provides suspects with specific rights which are not foreseen in other countries.
Currently, the Eurojust/Europol supported SIRIUS Project helps judicial and law enforcement authorities in this pursue, by creating a repository of applicable procedures and publishing a yearly report \cite{europol2021sirius} on the status of obtaining digital evidence from OSPs.

Several challenges have been highlighted in relation to the automation of communication and evidence sharing in the context of cross-border collaborations. Although a single tool that could enable the automation of all the required procedures would be desirable, the difficulties of building such a highly granular tool are daunting. Nevertheless, when approached individually, several technologies could ease the automation of such tasks. For instance, blockchain and smart contracts could provide enough guarantees to automate investigation requests, which would be digitally signed and audited. In the case of evidence collection, similar procedures could be used, exploiting the existing tools leveraged to collect evidence and enhancing them with the tamper-proof capabilities of blockchain, the use of hashes, and encryption. Furthermore, such tools could be linked with local jurisdictions systems through APIs. More details on the benefits of blockchain are given in Section \ref{sec:discussion}.

Data localisation policies have been extensively discussed in the literature as a strategy to reduce the burden of data acquisition both in terms of legal requests and the related technical issues \cite{abraha2021law,gppi2019}. For example, it is often unclear which jurisdiction determines the applicable procedural framework that regulates the gathering and validity of digital evidence as well as to which jurisdiction EIOs or MLAs should be sent \cite{europolchallenges,europol2021sirius}. However, data localisation has several drawbacks, such as threatening the privacy of individuals should data be stored in jurisdictions under the control of governments with weak human rights protections. The latter could also cause conflicts of law and hinder the resolution of cross-border investigations. Moreover, modifying the decentralised nature of such systems would affect their security, resiliency and performance. Last but not least, data localisation has economic factors preventing its practical application, such as limiting the exploitation capabilities of the involved organisations, which results in the organisations' reluctance to adopt data localisation policies \cite{Chauhan2021}. 

The access to that data by JA/LEA, inevitably triggers discussions about balancing the right to privacy and secrecy of communications with the need for ensuring public security and effectively tackling serious crime.
The Court of Justice of the European Union (CJEU) has taken the approach to set limits on data retention regimes and impose access conditions to retained data since 2014, when the CJEU declared the 2006 Data Retention Directive to be invalid. More recently, the Court has admitted exceptions to those rules \cite{siochana,cybmonitor,quadrature} and the possibilities for OSPs to retain data and for public authorities to use that data in criminal processes are part of an increasingly complex framework.

\section{Discussion - Enhancing Cross-border Collaboration}
\label{sec:discussion}

\subsection{Current Activities and Related Projects}

The European Commission has granted several projects aiming to provide solutions for cybercrime prevention and prosecution, and facilitate common procedures in the management of digital evidence. Thus, in this section we examine the extent to which the EU has promoted initiatives aimed at increasing the security level of the actors involved in the fight against cybercrime.

With the aim to provide solid results of the above-mentioned topic, we used the EU's CORDIS \cite{cor} database to perform our search as it stores the information and public deliverables of all EU-funded projects. The two searches performed consisted in:  

\begin{itemize}
\item Finding projects listed with fields ``criminology'' or ``law enforcement''. Therefore, the search consisted in \texttt{((criminology OR law enforcement) AND status == SIGNED)} in all projects' category with an all-years timespan from 1990 to 9th February 2022. The query, after eliminating duplicate entries, returned 72 projects. Projects not focused on digital crime, not providing frameworks to facilitate investigation procedures, or tools to empower law enforcement bodies where rejected by consensus. Thus, 23 references were accepted in the qualitative synthesis.
\item Finding projects granted under the H2020 framework with call identifiers ``H2020-FCT-2014-2015'', ``H2020-FCT-2016-2017'', ``H2020-SU-SEC-2018'', ``H2020-SU-SEC-2019'', and ``H2020-SU-SEC-2020''. From the 132 obtained results, projects found in the previous search or not focused on digital crime, investigation procedures, or tools aiming to improve law enforcement capabilities against cybercriminals where rejected by consensus. Consequently, 16 projects were finally accepted in the qualitative synthesis. 
\end{itemize}

The identified records, listed from the most recent starting date, are depicted in Table \ref{tab:related_projects}. The table provides the scientific field of the project, the project acronym, the starting, and the ending dates.

From the 39 projects included in the analysis after the screening, two main groups were identified:
\textit{Cross-Border Governance and Enforcement Services}, and \textit{Tool-kits Development for Law Enforcement}. We discuss each category next.

\subsubsection{Cross-Border Governance and Enforcement Services}
The projects classified under this category aim to provide frameworks and procedures to address the lack of common regulations and the several disparities associated with criminal prosecution. Also, these projects bring new tools to empower law enforcement and judiciary bodies in their fight against all forms of digital crimes. 

In an attempt to improve the protection of victims of human trafficking and child sexual abuse, HEROES and GRACE will provide technology to build bridges and facilitate cross-border coordination amongst law enforcement agencies, prosecutors, judges, and civil society organisations. In particular, GRACE will counter the spread of online child sexual exploitation material with the deployment of advanced analytical and investigative mechanisms. Results will be implemented by Europol and used by European LEAs. In this line, LOCARD will use Machine Learning (ML) algorithms to develop tools seeking for potential pedophile behaviours on social networks. Moreover, the platform provided by LOCARD aims to guarantee the integrity and transparency of the cross-jurisdictional chain of custody with blockchain technology. A similar objective is also shared by CREST that will implement the same technology to manage and deliver court-proof digital evidence. The project will deliver a platform to help LEAs fighting cybercrime in IoT ecosystems, autonomous systems, and targeted technologies. Not with blockchain but aiming to improve cross-border exchange of information, SHUTTLE will deploy a toolkit in accordance with the ISO17025 fostering, therefore, the use of a common methodology across European countries. Also, the increasing involvement of mobile phones in cybercrimes led the EU to invest in FORMOBILE and EXFILES projects. Whilst FORMOBILE will provide tools to facilitate investigations in mobile devices and develop a standard to homogenise forensic workflows, EXFILES will focus its efforts on the data extraction of encrypted files. Not focused on mobile devices but on providing a comprehensive picture of the presented evidence, the already ended project SPIRIT brought capabilities for LEAs across the investigation workflows and empowered them in criminal investigations. In particular, the project developed heterogeneous relationships on social graphs and provided privacy by design tools to enhance the acquisition and analysis phases of the investigation.

The improvement of law enforcement interactions using Artificial Intelligence (AI) tools represent a main objective for the EU project, pop AI. The project will increase trust in AI by building an ecosystem comprising several European LEAs. Similarly, ALIGNER and LAW-GAME will allow relevant European actors to identify and discuss needs to develop AI tools aiming to support, train and empower law enforcement bodies. From a different perspective but also aiming to foster dialogue amongst judiciary forces, and all actors involved in the investigation process, CYCLOPES and PROTAX will seek to connect with the most relevant European and international bodies, build bridges between industry and academia, and face the current challenges associated to fighting cybercrime (i.e., procedures, training, or standardisation). Whilst CYCLOPES is focusing its efforts on building and maintaining a network of LEAs, PROTAX has also involved tax authorities to improve the prevention and prosecution of tax crimes. In this line, the EU projects I-LEAD and ILEAnet stand for building solid and sustainable LEA networks focused on research and seeking for innovative needs. Although I-LEAD will provide recommendations to improve standardisation procedures, ILEAnet will establish a network of LEA practitioners to foster innovation and share best practices across the community. Also, the EU project COPKIT has developed a platform to address issues related to the investigative processes of LEAs (i.e., analysis, investigation, mitigation and prevention). A panel of end users and stakeholders (led by EUROPOL) aim to ensure the coherence of the results. Similarly, ROXANNE will provide tools with shared intelligent and speech processing technologies to help LEAs make decisions in situations of high-levels of pressure. Moreover, the shared intelligent platform developed by INSPECTr will provide a novel process to help law enforcement in predicting, detecting and managing crimes at national and supranational levels. Finally, ASGARD, a project ended in 2020, built a sustainable community (comprising LEAs and actors of the research industry) aiming to develop tools to extract, exchange and analyse large volumes of data for forensic investigations. Moreover, the project performed several actions to foster the interaction among stakeholders and enhance trust.

The disparities and complexity of new tools to enable a better management of digital investigations for LEAs, might increase the need for skilled experts to investigate cybercrime cases. Thus, the TUECS project will face this gap by providing stakeholders an innovative governance theory to foster public and private partnership while reducing their cooperation costs. Also, the high demand for security experts to fight cybercriminals has also been addressed by ESSENTIAL. With the implementation of a broad range of research topics, the project will provide effective security and interdisciplinary training campaigns to relevant experts and professionals. 

The international nature of most criminal cases fosters the need for developing regulations led by authorities located around the globe. Although JustSites is not purely focused on digital crimes, this EU project will study the most relevant locations of international criminal authorities contributing, therefore, to a better understanding of their role and influence in criminal prosecutions.

\subsubsection{Tool-kits Development for Law Enforcement}
The large number of tools used by LEAs and actors involved in judicial processes along with their technical limitations, frequently lead to inefficient investigation procedures. Thus, with the aim to build stronger, more resilient and effective prosecutions, the development of tools to help LEAs must address the above mentioned challenges and provide the latest technology and innovative features. The previous section was focused on projects aiming to develop tools and foster standardisation, participation, and coordination of European police forces across the Member States, whereas this section provides a summary of projects, granted by the EU Commission, seeking to undertake strong research programs and develop innovative tools with the most relevant technological advances and benefits for LEAs. 

With the participation of ten different LEAs, practitioners and combining several fields of expertise (i.e., technology, sociology, psychology, linguistics and data science), the PREVISION project provides a platform to face the most relevant cross-border security challenges. Likewise, the innovative research program, iCrime, will explore the different pathways of cybercrime offenders. The project aims to improve the understanding of cybercrime markets from social and economic perspectives. Using a different approach and after consulting the needs of several European LEAs, MAGNETO has developed sophisticated solutions and tools to address the lack of heterogeneity and other problems arising from the use of massive volumes of data.      

Based on the fact that cybercriminals are improving data hiding methods (e.g., steganography) to perpetrate their malicious activities, UNCOVER seeks to provide LEAs with tools committed to bridge the gaps left open by commercial solutions (e.g., limited number of hiding methods, slow performance, or lack of confidence). The solution will consider users' operational needs, regulations, and chain of custody considerations. Similarly, the already ended project RAMSES combined scraping techniques of public and deep web to bring the latest advances in an intelligent steganalysis software platform to detect manipulation in images and videos. The platform could detect and track malware payments and extract and analyse malware samples using Big Data algorithms. In this line and aiming to enhance the collaboration amongst the many actors involved in the prosecution of a crime, APPRAISE will bring together representatives from a wide range of disciplines (i.e., technology, psychology, sociology) and society to overcome the several complexities of fighting against cybercriminals. Also, aiming to foster data exchange and communication amongst LEAs, PROACTIVE will support the EU Action Plan for Chemical, Biological, Radiological and Nuclear (CBRN) threats by providing innovative tools to improve the response capacities of policymakers, security professionals and the civil society. Likewise, TRACE and AIDA will provide solutions to identify, track and document all actions performed in investigation workflows. Whilst TRACE is focused on illicit financial flows (IFFs), AIDA addresses cybercrime with tools using data mining and analytics solutions. Similarly, by combining augmented reality and machine learning algorithms, DARLENE and INFINITY will provide solutions to improve LEAs decision-making preventing, therefore, criminal activities. In this line, the EU projects CounteR and INDEED will develop tools and capabilities to counter radicalisation in Europe and encourage LEAs to undertake coordinated actions. Also, aiming to improve the understanding of the psychological dimension of cybercriminals, CC-Driver will perform a thorough research on human factors leading to all forms of cybercrime and will deliver tools to prevent, investigate and mitigate cybercriminal behaviour. Results will maximise potential victims' protection and contribute to more effective training campaigns.

The identification of perpetrators using DNA analysis has several limitations in forensic investigations. With the aim to address these challenges, VISAGE provides a toolkit with intelligence information on appearance, age, and ancestry to construct composite sketches (of unknown trace donors) from traces recovered at crime scenes. A set of tools was also deployed by VICTORIA that developed a Video Analysis Platform (VAP) to address the lack of maturity related to video investigation tools.  

\textit{``51\% of EU citizens feel not at all or not well informed about cyber threats and 86\% of Europeans believe that the risk of becoming a victim of cybercrime is rapidly increasing.''} This conclusion was highlighted in RAYUELA \cite{ray}, a project seeking to educate young people in the use of the Internet, therefore, preventing and mitigating cybercriminal behaviour.

\subsubsection{Other Relevant Projects}
Projects described in the previous section have been extracted from CORDIS. This database provides a comprehensive and structured public repository with all the information on projects whose funding, totally or partially, comes from the European Commission. However, there are other initiatives promoted by European institutions that, due to their relevance and potential impact, are worth mentioning.

With the participation of Europol, Eurojust, and the European Judicial Network, the SIRIUS project \cite{sir} aims to provide guidelines on specific Online Service Providers (OSPs) along with investigative and analytical tools developed by Europol and the Member States. Moreover, the project would facilitate exchange of information and experience sharing amongst all parties involved in cybercrime prosecutions. In parallel, the Council of Europe is also playing a relevant role in the fight against cybercrime. Besides the already mentioned Cybercrime Convention, which the CoE has been developing and promoting through the Octopus Project \cite{oct} over the years, the institution is also promoting other initiatives like GLACY+, iPROCEEDS-2, CyberSouth or CyberEast seeking to improve cybercrime investigations in the international arena.

Funded by the European Commission, the European Cybercrime Training Education Group (ECTEG) \cite{ect} provides training and education material to build law enforcement capacity on issues related to cybercrime. Amongst its most relevant projects, the Global Cybercrime Certification Project (GCC) seeks to create a common, international and harmonised certification system for law enforcement agents and judiciary forces. Likewise, the DECRYPT project improves law enforcement continuous education by providing e-learning and classroom materials aimed at addressing encryption issues for decrypted materials to be admitted in a court of justice.

\begin{table*}[hbt!]
\renewcommand{\arraystretch}{1.1}
 \centering
  \caption{Projects granted by the EU, ordered by most recently started.}
 \scriptsize
 \rowcolors{2}{gray!25}{white}
 \begin{tabular}{p{3.25in}p{1in}p{0.75in}p{0.75in}}
\toprule 
\textbf{Fields of science} & \textbf{Project acronym} & \textbf{Start date} & \textbf{End date} \\
 \midrule
civil society; criminology; human trafficking; law enforcement & HEROES \cite{her} & 01/12/2021 & 30/11/2024 \\
ecosystems; civil society; artificial intelligence; ethical principles; law enforcement & pop AI \cite{pop} & 01/10/2021 & 30/09/2023 \\
civil society; artificial intelligence; law enforcement & ALIGNER \cite{ali} & 01/10/2021 & 30/09/2024 \\
public policies; law enforcement; ideologies & INDEED \cite{inde} & 01/09/2021 & 31/08/2024 \\
virtual reality; law enforcement; & LAW-GAME \cite{law} & 01/09/2021 & 31/08/2024 \\
artificial intelligence; law enforcement; big data & APPRAISE \cite{app} & 01/09/2021 & 31/08/2023 \\
criminology; computer and information sciences; law enforcement & iCrime \cite{icr} & 01/07/2021 & 30/06/2026 \\
monetary and finances; law enforcement & TRACE \cite{tra} & 01/07/2021 & 30/06/2024 \\
criminology; law enforcement & UNCOVER \cite{unc} & 01/05/2021 & 30/04/2024 \\  
network security; law enforcement & CYCLOPES \cite{cyc} & 01/05/2021 & 30/04/2026 \\ 
data protection; social psychology; law enforcement; data mining & CounteR \cite{cou} & 01/05/2021 & 30/04/2024 \\
ergonomics; law enforcement; Internet & RAYUELA \cite{ray} & 01/10/2020 & 30/09/2023 \\
ecosystems; Internet of Things; law enforcement & DARLENE \cite{dar} & 01/09/2020 & 31/08/2023 \\
law enforcement; data mining; terrorism; big data; deep learning & AIDA \cite{aid} & 01/09/2020 & 28/02/2023 \\
software; criminology; mobile phones; law enforcement & EXFILES \cite{exfi} & 01/07/2020 & 30/06/2023 \\
eCommerce & GRACE \cite{grac} & 01/06/2020 & 30/11/2023 \\
artificial intelligence; law enforcement; big data & INFINITY \cite{infi} & 01/06/2020 & 31/05/2023 \\
governance; forensic sciences; law enforcement & TUECS \cite{tue} & 01/06/2020 & 31/08/2022 \\
ergonomics; criminology & CC-DRIVER \cite{ccd} & 01/05/2020 & 30/04/2023 \\
law enforcement; big data & INSPECTr \cite{insp} & 01/09/2019 & 28/02/2023 \\
data protection; criminology; phonetics; law enforcement; natural language processing & ROXANNE \cite{rox} & 01/09/2019 & 31/12/2022 \\
criminology; big data & PREVISION \cite{pre} & 01/09/2019 & 31/12/2021 \\
ergonomics; ecosystems; law enforcement; terrorism & CREST \cite{cre} & 01/09/2019 & 28/02/2023 \\
criminology; electrical engineering; mobile phones; forensic sciences; law enforcement & FORMOBILE \cite{form} & 01/05/2019 & 30/04/2022 \\
criminology; & LOCARD \cite{loc} & 01/05/2019 & 31/07/2022 \\
civil society; law enforcement & PROACTIVE \cite{pro} & 01/05/2019 & 30/04/2022 \\
planetary geology; criminology & JustSites \cite{jus} & 01/01/2019 & 31/12/2023 \\
data protection; active learning; computational intelligence; law enforcement & SPIRIT \cite{spi} & 01/08/2018 & 31/10/2021 \\
ecosystems; ethical principles & COPKIT \cite{cop} & 01/06/2018 & 30/09/2021 \\
software; databases; bayesian statistics; colors & SHUTTLE \cite{shut} & 01/05/2018 & 30/04/2022 \\
machine learning; virtual reality; criminology; ontology; law enforcement; data mining; terrorism & MAGNETO \cite{magn} & 01/05/2018 & 30/04/2021 \\
data protection; ergonomics; taxation; criminology; law enforcement & PROTAX \cite{protax} & 01/05/2018 & 31/07/2021 \\
law enforcement & I-LEAD \cite{ile} & 01/09/2017 & 28/02/2023 \\
law enforcement & ILEAnet \cite{ilea} & 01/06/2017 & 31/05/2022 \\
DNA; software; criminology; colors & VISAGE \cite{vis} & 01/05/2017 & 31/10/2021 \\  
data protection; mobile phones; optical sensors; computer vision; law enforcement & VICTORIA \cite{vict} & 01/05/2017 & 30/11/2020 \\  
law enforcement & ESSENTIAL \cite{ess} & 01/01/2017 & 31/12/2021 \\ 
radio and television; law enforcement; data mining; big data & ASGARD \cite{asga} & 01/09/2016 & 30/11/2020 \\
malicious software; criminology; forensic sciences; law enforcement; internet & RAMSES \cite{rams} & 01/09/2016 & 30/11/2019 \\

 \bottomrule
 \end{tabular}

 \scriptsize
 \label{tab:related_projects}%
\end{table*}

\subsection{Commercial solutions - Existing Tools }

\begin{table*}[th!]
\renewcommand{\arraystretch}{1.1}
 \centering
 \caption{Relationship between the challenges and the DEMS' evaluated features. }
 \scriptsize
 \rowcolors{2}{gray!25}{white}
 \begin{tabular}{p{2.3in}ccccc}
\toprule 
\textbf{Challenge} & \textbf{Evidence} & \textbf{Reporting} & \textbf{Chain of custody} & \textbf{Use of} & \textbf{Regulations} \\
& \textbf{collection} & \textbf{tools} & \textbf{assurance} & \textbf{standards} & \textbf{compliance} \\
 \midrule
 Individual's control over their own data & & & & & $\checkmark$ \\
  Timely collection and sharing of evidence & $\checkmark$ & & & & \\
  Lack of harmonisation in rules of admissibility of criminal evidence and prosecution & & $\checkmark$ & $\checkmark$ & $\checkmark$ & \\
  Lack of compatibility between different protocols regarding data categorisation and definitions & & & & $\checkmark$ & \\
  Direct cooperation with service providers and equality of opportunities  & & & & & $\checkmark$ \\
  Incompatibility conflicts between jurisdictions that may violate procedural laws and rights & & & & $\checkmark$ & $\checkmark$ \\
Lack of automated mechanisms to efficiently collect and report requests & $\checkmark$ & & $\checkmark$ & & \\
  Auditability in data collection procedures & &$\checkmark$ & & $\checkmark$ & $\checkmark$ \\
  Lack of resources related to equipment and training of law enforcement and judicial authorities to support direct co-operation between different jurisdictions & & & &$\checkmark$ & $\checkmark$\\
    Data retention issues &$\checkmark$ & &$\checkmark$ & &$\checkmark$ \\
 \bottomrule
 \end{tabular}
 
 \scriptsize
 \label{tab:relationship_tools_challenges}%
\end{table*}

Digital evidence management systems (DEMS) are the main commercial solutions to manage digital forensic investigations. In what follows, the most relevant DEMS available in the market are analysed and compared. To this end, several features of such DEMS have been considered, namely the mechanism to collect digital evidence, the reporting tools, the assurance of the chain of custody, the use of standards, and the compliance with regulations. Table \ref{tab:relationship_tools_challenges} provides the relationship between the challenges and these features. The comparison of 34 DEMS is summarised in Table \ref{tab:market}.
Concerning digital evidence collection, it is traditionally conducted manually by the investigator in charge of the criminal investigation. However, to shorten investigation times and optimise resources, further more automated procedures have already been considered, such as the use of public portals where citizens can upload potentially valuable resources for ongoing investigations, and the automatic collection of evidence by scanning the data stored in devices directly connected to the DEMS. Whereas manual procedures are implemented in all solutions, automated procedures are considered in only six (18\%). 
The addition of automated mechanisms is a must in future solutions.
Another popular feature of DEMS is the ability to create reports summarising, among others, the insights acquired from the investigations to be presented in court, or the audit trails with the chronological set of records related to the investigations and their digital evidence. These tools are crucial to provide accountability for the entire investigation procedures, demonstrating that they have been conducted in a lawful, transparent and trustworthy way. Surprisingly, a fourth of the analysed DEMS do not provide any reporting functionality (9/34, 26\%). 62\% of the DEMS (21/34) enable audit trails reports, and 29\% (10/34) enable exporting court-accepted reports as part of the documentation related to the digital investigations. With regards to these court reports, DEMS do not mention in which jurisdictions or courts of justice are these reports accepted, a very valuable information due to the disparity and discrepancies among jurisdictions. Standardising and harmonising court reports will gain significant relevance in the incoming years due to the increase of cross-border crimes.

\begin{table*}[th!]
\renewcommand{\arraystretch}{1.1}
 \centering
  \caption{Analysis and comparison of commercial tools.}
 \scriptsize
 \rowcolors{2}{gray!25}{white}
 \begin{tabular}{p{0.8in}p{1.5in}p{0.65in}p{0.95in}p{0.9in}p{1.1in}}
\toprule 
\textbf{Commercial solution} & \textbf{Evidence collection} & \textbf{Reporting tools} & \textbf{Chain of custody assurance} & \textbf{Use of standards} & \textbf{Regulations compliance} \\
 \midrule
  ADF \cite{adf} & Manual & Court reports & N/A & N/A & N/A \\
  ARQ \cite{arq} & Manual, compatible devices & Audit trails & Hash (SHA-256) & N/A & N/A \\
  AXO \cite{axo} & Manual & Audit trails, court reports & Yes* & N/A & CJIS \\
  CEL \cite{cel} & Manual & No & Yes* & N/A & N/A \\
  CCE \cite{cce} & Manual, compatible devices & No & Yes* & N/A & N/A \\
  DET \cite{det} & Manual & Audit trails & Yes* & N/A & N/A \\
  DOT \cite{dot} & Manual & Audit trails & Yes* & N/A & N/A \\
  DTQ \cite{dtq} & Manual & Court reports & Hash* & N/A & N/A \\
  DOQ \cite{doq} & Manual & Audit trails & Yes* & SWGIT & N/A \\
  ECF \cite{ecf} & Manual, compatible devices & Court reports & N/A & N/A & N/A \\
  ERI \cite{eri} & Manual & Audit trails, court reports & Yes* & N/A & N/A \\
  EVW \cite{evw} & Manual & No & Yes* & N/A & MoPI \\
  FOR \cite{for} & Manual & No & Yes* & FedRAMP & CJIS \\
  GEN \cite{gen} & Manual & No & N/A & N/A & N/A \\
  HIT \cite{hit} & Manual & Audit trails & Yes* & N/A & MoPI, GDPR \\
  HYT \cite{hyt} & Manual & No & N/A & N/A & N/A \\
  INS \cite{ins} & Manual & Court reports & N/A & N/A & N/A \\
  KIN \cite{kin} & Manual & Audit trails & Yes* & N/A & Yes* \\
  LIM \cite{lim} & Manual & Audit trails & Yes & ISO & N/A \\
  LIN \cite{lin} & Manual & No & Yes & N/A & N/A \\
  NEW \cite{new} & Manual & Audit trails & N/A & FIPS 140-2 & CJIS \\
  NIC \cite{nic} & Manual, public portal & Audit trails & Yes & N/A & CJIS \\
  OMN \cite{omn} & Manual & Audit trails, court reports & Yes & N/A & Yes \\
  ORA \cite{ora} & Manual & Audit trails, court reports & Yes & N/A & CJIS \\
  PAT \cite{pat} & Manual & Audit trails & Yes & N/A & N/A \\
  PWI \cite{pwi} & Manual & Audit trails & N/A & N/A & N/A \\
  SAF \cite{saf} & Manual & No & Yes & N/A & N/A \\
  SFL \cite{sfl} & Manual & Audit trails & Yes & N/A & CJIS \\
  SPD \cite{spd} & Manual, public portal & No & Hash (SHA-256) & N/A & CJIS, CDR, IRS, DoD \\
  UDE \cite{ude} & Manual, compatible devices & Audit trails, court reports & Hash* & N/A & N/A \\
  VER \cite{ver} & Manual & Audit trails & Hash (patented) & FIPS & CJIS \\
  VDZ \cite{vdz} & Manual & Audit trails, court reports & Hash (SHA-256) & FedRAMP, FIPS 140-2 & CJIS, HIPAA, GDPR, DoD, ITARM, EAR\\
  WOL \cite{wol} & Manual & Audit trails & Hash (SHA-256) & N/A & CJIS \\
  XWI \cite{xwi} & Manual & Audit trails & N/A & N/A & N/A \\
  \multicolumn{6}{l}{* No further details provided online.}  \\
 \bottomrule
 \end{tabular}
 \label{tab:market}
\end{table*}

Ensuring the chain of custody of digital evidence is another critical feature. To achieve successful prosecution, the integrity of evidence needs to be guaranteed and proved, from their initial gathering to their final presentation in court. Hence, tamper-proof solutions are required. In general, the cryptographic solutions to ensure the integrity of any file are one-way hashing functions. In case of tampering (intentionally or accidentally) a digital evidence, the resulting hash will be different and, in consequence, the chain of custody broken.
Despite its importance, the majority of the analysed DEMS do not provide many details about this fundamental feature. For instance, 19 tools (56\%) mention that the chain of custody is guaranteed, but no further details about how this is achieved are provided, whilst 8 tools (23\%) do not mention this feature at all.
The other 7 solutions explicitly mention the use of hashing mechanisms. More specifically, four of these solutions (ARQ, SPD, VDZ and WOL) use the well-known SHA-256 algorithm, and the VER tool uses a US patented interlocking hashing.
However, the management of these hashes to ensure the evidence chain of custody is not detailed. Future solutions should clearly describe the technologies and processes involved in the assurance of the chain of custody.

In order to bring digital evidence to the courts of law, it is necessary to follow the national standards, laws and methodologies regarding the chain of custody. Unfortunately, international standards for digital investigations are not common, despite the many extant guidelines and documents from national organisations and LEAs.
The lack of standards is reflected in the number of DEMS adopting them. Indeed, only 6 tools (18\%) use some standard.
More specifically, the US FIPS\footnote{Federal Information Processing Standard} standard is adopted by NEW, VER and VDZ; the US FedRAMP\footnote{Federal Risk and Authorization Management Program} standard is adopted by FOR and VDZ; the SWGIT\footnote{Scientific Working Group on Imaging Technology} standard is adopted by DOQ; and quality standards set by the ISO are adopted by LIM.

Assessing the impact of the DEMS in terms of social/ethical responsibility, fundamental rights, data protection and privacy is mandatory to stand by the current regulations and legislations. Generally, the GDPR has harmonised the data protection laws across EU member states by strengthening data processing principles and granting citizens with extensive rights. Regulations intended for LEAs and national security/intelligence parties are, among others, the FBI’s CJIS\footnote{Criminal Justice Information Services} in the US, or the MoPI\footnote{Management of Police Information} in the UK.
Surprisingly, only 14 DEMS (41\%) recognise that they comply with some regulation.
For instance, 10 tools comply with the CJIS regulation since their market is mostly located in the US. However, only two DEMS (HIT and VDZ) are GDPR-compliant. Similarly, MoPI-compliant DEMS are only EVW and HIT. Other regulations implemented in DEMS are the US DoD\footnote{Department of Defence} regulatory program in SPD and VDZ; and the US HIPAA\footnote{Health Insurance Portability and Accountability Act}, EAR\footnote{Export Administration Regulations} and ITARM\footnote{International Traffic in Arms Regulations} in VDZ.

\subsection{The Road Ahead}
Inevitably, due to the need for exchanging digital evidence there will appear more initiatives, beyond the aforementioned. One of the key elements in this discussion is the \emph{chain of custody} as we are considering cases which are initiated in a jurisdiction and are followed up in another with the control handed over from an entity to another, partially or as a whole. An obvious choice would be to determine whether the control might be centralised or decentralised. We sustain that the decentralised option is more appealing as it allows for more flexibility and control in each jurisdiction and prevents the issues of single points of failure. Moreover, with the introduction of blockchains and distributed ledgers there are several issues that can be inherently tackled, e.g. traceability, auditability, and, of course, immutability. Notably, the use of smart contracts can facilitate the automation of such exchanges and enable fine-grained control of who has access, when, what can be submitted and exchanged, by whom etc. The latter introduces other practical issues as, for instance, existing legislation does not allow LEAs to use platforms and store evidence in public facing storage facilities or use infrastructure that common civilians use, let alone civilians from different countries.

The creation of dedicated platforms, such as the eEDES and others, based on blockchain technology must be streamlined in such a way that the role of national judicial and law enforcement authorities is correctly balanced with the intervention of supra national entities, such as Eurojust and Europol. The possibility for a central authority to intervene in such platforms should only be included if they are designed for cooperation with non EU States \cite{institutional_espina}. 
Technical and legal solutions designed to deal with digital evidence, often very volatile, need to find fast and direct routes for information gathering and sharing and not shy away from the inclusion of public/private cooperation, as direct cooperation among two sectors is often a necessary strategy to fight crime in the digital age. In this direction goes the latest decision to empower Europol \cite{europol_mandate}, with the appropriate supervision, and allow it to process large datasets and receive data from private companies.

Setting aside issues such as identity management and access rights, which are more technical, an important aspect that has to be considered is the admissibility of digital evidence in court. The questions that emerge are primarily related to the collection of digital evidence. For instance, the collection of digital evidence by involving specific methods might be admissible in one country but not in another. Thus, the exchange of digital evidence would be legal, but the evidence would not be admissible. This is rather important especially in the eye of authoritarian regimes, lawful interception, deception during interrogations, and use of AI and machine learning against use of, e.g. decentralised platforms and end-to-end encryption. All the above, individually, may punch holes in the admissibility of evidence in court while raising ethical issues. The case of using the notorious Pegasus spyware \cite{pegasus} while exceptional, clearly illustrates how different countries consider lawful interception and surveillance. Moreover, the legality of using specific tools, methods, and the overall practice of the judicial system is questionable in many authoritarian regimes and may result in further violations of human rights. 

Of particular interest is the recurring discussion on encryption and access to the underlying data from the LEAs. Clearly, the abuse of encryption by criminals, not only cyber criminals, introduces many additional burdens for LEAs and digital forensics experts. This is something that troubles law and policy makers \cite{council_enc} regardless of the laws that have been adopted  \cite{tola,uk_act} or plan to be adopted by some countries \cite{uk_draft}, especially targeting end-to-end encryption. The red line between excessive surveillance capabilities and providing LEAs with the necessary access can be very thin. Even more, measures to prevent unintended negative side effects might not be enforceable as the integration of a backdoor in an encryption algorithm practically renders encryption useless and jeopardises the protection of fundamental rights and citizens' data. The above introduces more questions regarding who is collecting the digital evidence, how, and whether this collection is acceptable to the rest of the parties in the chain of custody of a case.

Finally, we sustain that once the legal and ethical aspects are tackled, standardisation activities should allow for the technical development of such solutions in an operational manner. Standardisation should cover the definition of entities, roles, underlying ontologies, and the allowed interactions among entities. 

There is still a long road ahead to achieve the proper alignment between the required protocols enabling cross-border prosecution, the underlying evidence management systems from a practical perspective, and other legal, ethical and procedural aspects that are continuously evolving to be on track with the current state of practice. In this regard, we sustain that novel directives and initiatives, such as the AI Act \cite{aiact} should take into account the challenges and views discussed in this article to avoid introducing more burden to current challenges while trying to solve others. The latter is crucial, especially in the case of AI and machine learning, since they are continuously being integrated into many software solutions and are used by LEAs and digital investigators. Therefore, we think that more communication and collaboration is needed between policy makers, LEAs, digital investigators, academic and legal experts, as well as representatives of the general public, to reach to solutions conforming to the current ethical values and respecting the freedoms and rights of individuals to fight against next-generation cybercrime.

\section{Conclusions}
\label{sec:conclusions}

The sophistication of criminal activities paired with ICT evolution hinder current investigations and require continuous cross-border collaboration between different entities. The latter is not an easy task since several challenges arise, e.g., in the legal, technical, and ethical dimensions. The research questions posed in Section \ref{sec:methodology} summarise the main aim of our research, namely providing a comprehensive state of knowledge of the different mechanisms to leverage cross-border investigations, their challenges, and a fruitful discussion of the road ahead of this particular matter. We discuss them in order next:

\begin{quote}
    \textit{\textbf{Q1:} Which are the current tools, procedures, and protocols for cross-border evidence exchange amongst European countries/jurisdictions?}
\end{quote}

In order to provide enough background to discuss the rest of the research questions, we have summarised the main mechanisms and protocols enabling cross-border collaboration. In Section \ref{sec:background} we have provided this information in the context of Europe, along with other well-known international procedures. According to our analysis, each mechanism has a different scope, and the application of the proper one is required in each case, especially to minimise the investigation's overhead.

\begin{quote}
    \textit{\textbf{Q2:} Which are the main challenges related to cross-border investigations?}
\end{quote}

A profound analysis of the literature was required to extract all the challenges of this particular matter, as described in Section \ref{sec:methodology}. The selection of articles and reports that discussed the current challenges of cross-border investigations allowed us to conclude that the same issues are identified by different authors regardless of their background. We have summarised and abstracted these challenges to provide a clear overview of the state of practice, and we have discussed them along with some possible countermeasures in Section \ref{sec:challenges}. 

\begin{quote}
    \textit{\textbf{Q3:} Are current practices efficient enough to counter the sophistication of cybercrime?}
\end{quote}

To answer this research question, we need to combine the information from the two previous ones. In a nutshell, the current mechanisms used for cross-border collaboration are solving partial issues and challenges, but there is no panacea. Moreover, some recent mechanisms and protocols solve some of the identified challenges while introducing new ones, despite the efforts of the actors involved in the process. Thus, the outcome of this analysis is that a profound discussion is required amongst all stakeholders, followed by fast and efficient actions, since cybercriminals seem to be ahead of current legislation.

\begin{quote}
    \textit{\textbf{Q4:} What technologies or strategies can be used to deal with the identified challenges?}
\end{quote}

As summarised in Section \ref{sec:discussion}, continuous efforts are being made to ease cross-border investigations in terms of tools, technologies, research projects and legislation updates. However, there is still a long road ahead as current solutions are not sufficient to solve the existing challenges. With this aim, we set the ground for the next steps that should be tackled, along with some strategies highlighting the most urgent issues to be solved, which are creating bottlenecks and preventing efficient and robust prosecution. Moreover, we have discussed other possible issues that may arise in the near future, either standalone or due to a combination of challenges, so that prevention mechanisms can be put in place accordingly.

We think that the information analysed and the research questions answered in this article reflect the current state of practice with high fidelity. Therefore, this article provides a fruitful and interdisciplinary ground of research and a clear overview of the measures that may need to be considered in the years to come. 

As a final note, we sustain that enabling technologies such as blockchain could enhance the auditability and transparency of several procedures performed during investigations. Several proposals that prove the capabilities of such a technology in the context of forensic investigations have been provided in the literature \cite{dasaklis2021sok,Kumar2021,lone2019forensic,zarpala2021blockchain}.  Moreover, blockchain could be used to automate several of the previously discussed procedures (e.g., evidence exchange). The latter could improve trust in legal systems and reduce the delays in investigations  \cite{Chauhan2021}. Of course, storing the evidence on the blockchain would not be the best option, e.g. consider the case of the evidence being a hard drive of some terabytes, however, off-chain mechanisms such as the IPFS \cite{benet2014ipfs} could efficiently fill in this gap.

\section*{Acknowledgements}
This work was supported by the European Commission under the Horizon 2020 Programme (H2020), as part of the projects  \textit{LOCARD} (\url{https://locard.eu}) (Grant Agreement no. 832735) and HEROES (\url{https://heroes-fct.eu/}) (Grant Agreement no. 101021801). F. Casino was supported by the Beatriu de Pinós programme of the Government of Catalonia (Grant No. 2020 BP 00035).

\section*{Conflict of Interest}
The authors reported no potential conflict of interest.

\bibliographystyle{plain} 

\bibliography{refs}

\end{document}